\documentstyle[pre,aps,multicol,epsfig]{revtex}

\begin{document}

\title{Domain Walls of Single-Component Bose-Einstein Condensates 
in External Potentials}

\author{P.G. Kevrekidis$^{1}$, B.A. Malomed$^{2}$, D.J.\ Frantzeskakis$^{3}$, 
A.R. Bishop$^{4}$, H.E. Nistazakis,$^{3,5}$ and 
R. Carretero-Gonz\'{a}lez$^{6}$ }

\address{$^{1}$ Department of Mathematics and Statistics, University of\\
Massachusetts,
Amherst MA 01003-4515, USA\\
$^{2}$ Department of Interdisciplinary Studies, 
Faculty of Engineering, Tel Aviv University,
Tel Aviv 69978, Israel \\
$^{3}$ Department of Physics, University of Athens,
Panepistimiopolis, Zografos, Athens 15784, Greece \\
$^{4}$ Center for Nonlinear Studies and Theoretical Division, Los Alamos
National Laboratory, 
Los Alamos, NM 87545, USA \\
$^{5}$ Department of Statistics and Actuarial Science, University of the
Aegean, Karlovassi, 83200 Samos, Greece \\
$^{6}$ Nonlinear Dynamical Systems Group,
Department of Mathematics and Statistics,
San Diego State University, San Diego, 
CA 92182-7720, USA; http://nlds.sdsu.edu/}

\maketitle

\begin{abstract}
We demonstrate the possibility of creating domain walls described by a single
component
Gross-Pitaevskii equation with attractive interaction, in
the presence of an optical-lattice potential. While it is found that the
extended domain wall is unstable, we show that the external magnetic trap
can stabilize it. Stable solutions include twisted domain walls,
as well as asymmetric solitons. The results also apply to spatial solitons
in planar waveguides with transverse modulation of the refractive index.
\end{abstract}



\section{Introduction}

Solitary waves, such as bright and dark solitons, are ubiquitous dynamical
patterns in the fields of nonlinear optics and matter waves \cite{b24,b25}.
Besides these customary waveforms, solitary coherent structures in the form
of domain walls (DWs) were also predicted in optical fibers with normal
group-velocity dispersion carrying two waves with orthogonal polarizations,
circular or linear, that interact through the cross-phase modulation induced
by the Kerr nonlinearity \cite{Wab,Malomed}. Similar structures were also
predicted in planar nonlinear optical waveguides \cite{Haelt}. The
fiber-optic DWs are distinguished by the property that, asymptotically, they
contain a single polarization, with a switch between the polarizations in a
localized region. Optical-DW solutions were, in fact, found following the
pattern of earlier known solutions of the DW (alias \textquotedblleft
grain-boundary\textquotedblright ) type in systems of coupled
Ginzburg-Landau equations that describe interactions between roll patterns 
\cite{MNT,Alik} or traveling waves \cite{Sakaguchi} with different
orientations in a convection layer .

Domain walls may also arise in very different physical media (that, however,
bear a similar mathematical description), namely in binary (two-component)
Bose-Einstein condensates (BECs). In particular, stable DW configurations
have been predicted in quasi-one-dimensional (cigar-shaped) binary BECs 
\cite{BECdw}. Such stable solutions can also exist in the presence of a periodic
optical-lattice (OL) potential \cite{opticalLattice}, combined with the
parabolic magnetic trap \cite{epjd}.

Optical DWs, as well as high-repetition periodic DW trains, have been
experimentally observed in fibers \cite{DWexperiment,DWarray}. On the other
hand, binary BECs have been experimentally achieved, for instance, in
mixtures of different spin states of $^{87}$Rb \cite{myatt}, including cases
when an OL was used for the confinement \cite{dsh}. Efforts were also made
to create binary BECs with different atomic species, such as $^{41}$K and 
$^{87}$Rb \cite{KRb}, or $^{7}$Li and $^{133}$Cs \cite{LiCs}. However, DWs
have not yet been observed, therefore a relevant issue is to predict more
favorable conditions for their existence.

Here, we propose a setting in which patterns of the DW type 
can be realized in a 
{\em single-component} BEC with the {\em attractive} interaction between
atoms. We demonstrate that this may occur, under suitable conditions, in the
presence of the OL potential. Furthermore, we show that the corresponding DW
solutions can be stabilized by the magnetic-trap potential. Our results also
predict the existence of patterns of the same type in nonlinear-optical
media, such as a multichannel nonlinear planar waveguide with transverse
modulation of the refractive index \cite{Wang}, or, possibly, in
two-dimensional photonic crystals. In particular, the attractive character
of the cubic nonlinearity, which is necessary for the existence of DWs, is
provided in optics by the usual Kerr effect. 

The presentation is structured as follows: in section II, we give a detailed
formulation of the model, and develop an analytical approach to the DW
solutions, based on the variational approximation (VA). In section III, we
construct the DW solution numerically, both in the absence and in the
presence of the magnetic trap; we obtain a variety of new solutions,
including an asymmetric DW-type soliton and a twisted DW, as well as
multi-DW patterns. Section IV concludes the paper.

\section{Formulation of the model and variational approximation}

Assuming that the nonlinear interactions are weak relative to tight
confinement in transverse dimensions, the transverse size of the
cigar-shaped condensate is much smaller than its length. In such a case,
which can be realized in strongly anisotropic traps, the Gross-Pitaevskii
(GP) equation, which governs the BEC in the mean-field approximation, assumes
an effectively one-dimensional form \cite{GPE1d}: 
\begin{equation}
i\psi _{t}=-\psi _{xx}+g|\psi |^{2}\psi +V(x)\psi ,  \label{dweq1}
\end{equation}
where $\psi $ is the single-atom wave function, $t$ and $x$ are measured,
respectively, in units of $2/\omega _{\perp }$ and the transverse
harmonic-oscillator length $a_{\perp }\equiv \sqrt{\hbar /(m\omega _{\perp })}$ 
($m$ and $\omega _{\perp }$ are the mass and transverse confining
frequency), and the energy unit is $\hbar \omega _{\perp }/2$. The
nonlinearity prefactor $g\equiv 2(2\pi )^{3/2}a/a_{\perp }$ is proportional
to the scattering length $a$ of the inter-atomic interactions (see e.g.,
Refs. \cite{GPE1d,review}), $a>0/a<0$ corresponding to the
repulsive/attractive interactions. In the general case, the potential is 
\begin{equation}
V(x)=\Omega ^{2}x^{2}+V_{0}\cos (kx),  \label{pot}
\end{equation}
where the two terms represent the magnetic trap and the OL, respectively. In
Eq.\ (\ref{pot}), $\Omega \equiv \omega _{x}/\omega _{\perp }$ ($\omega _{x}$
being the axial confining frequency) is the effective strength of the
magnetic trap, $V_{0}\equiv 2E_{{\rm rec}}/\hbar \omega _{\perp }$ is the OL
strength ($E_{{\rm rec}}\equiv h^{2}/2m\lambda _{{\rm laser}}^{2}$ is the
recoil energy, $\lambda _{{\rm laser}}$ being the wavelength of the
counter-propagating laser beams which generate the OL), and $k$ is the
wavenumber of the OL. In the experiment, it can be controlled not only by
changing $\lambda _{{\rm laser}}$, but also, more conveniently, by varying
the angle $\theta $ between the laser beams, as the the local intensity in
the interference pattern is modulated at the wavelength $\lambda \equiv 2\pi
/k=\left( \lambda _{{\rm laser}}/2\right) \,\sin (\theta /2)$ 
\cite{Morsch-Arimondo}.

To estimate actual physical quantities in this situation, we can take a
typical example of the attractive ($a<0$) condensate of $^{7}$Li containing 
$10^{3}$ atoms, confined in a cigar-shaped trap with the frequencies $\omega
_{x}=2\pi \times 60$ Hz and $\omega _{\perp }=2\pi \times 400$ Hz. This
implies $\Omega =0.15$ in Eq.\ (\ref{pot}), while the time and space units
correspond to $0.8$ ms and $2\mu $m, respectively. Finally, concerning the
OL wavenumber, $k=2$ (this is a value that will be dealt with in this work)
corresponds to the wavelength $\lambda \simeq 6$ $\mu $m of the interference
pattern (in physical units). These values can be used for the interpretation
of the results that are presented below in terms of the dimensionless
variables.

Stationary solutions to Eq. (\ref{dweq1}) are seeked for as $\psi
(x,t)=e^{-i\mu t}u(t)$, where $\mu $ is the chemical potential (proportional
to the energy per atom), and the real function $u$ obeys the equation
\begin{equation}
\mu u+u^{\prime \prime }-gu^{3}-V(x)u=0,  \label{dweq8}
\end{equation}
the prime standing for $d/dx$. DW-like solutions are those following the
pattern of the {\it ansatz}
\begin{equation}
u(x)=\left( A/2\right) \left[ 1-{\rm sgn}\left( x-\xi \right) \right] ,
\label{dweq2}
\end{equation}
where $A$ is the amplitude of the wave function at $x=-\infty $ (assuming an
infinitely long system), and $x=\xi $ is the location of the DW's center. In
linear quantum mechanics, a state of this type cannot exist because the
atoms would tunnel from the filled domain, $x<\xi $, into the empty one, 
$x>\xi $. However, in the GP equation (\ref{dweq1}), the nonlinear mean-field
term adds a negative contribution to the effective potential in the filled
domain. If, as a result, a state of the type (\ref{dweq2}) is possible with 
$\mu <0$, the tunneling will be suppressed, and a DW state may be supported.

Of course, such qualitative considerations provide 
no guarantee that a DW waveform will
really exist. To investigate this possibility in quantitative terms, we will
first resort to the VA (variational approximation), which will be followed
by direct numerical analysis in the next section. To this end, we note that
Eq. (\ref{dweq8}) can be derived from the Lagrangian, $L=\int_{-\infty
}^{+\infty }{\cal L}dx$, with the density 
\begin{equation}
{\cal L}=u_{x}^{2}-\mu u^{2}+V(x)u^{2}+(g/2)u^{4};  \label{dweq3}
\end{equation}
in this part of the work, the potential $V(x)$ does not include the
parabolic trap, see Eq. (\ref{pot}).

With the use of the ansatz (\ref{dweq2}), the integral which gives the
Lagrangian $L$ diverges as $x\rightarrow -\infty $. Isolating the diverging
part, and subjecting it to the variation in $A$, immediately yields a result 
\begin{equation}
A^{2}=\mu /g,  \label{dweq5}
\end{equation}
hence $\mu $ and $g$ should be of the same sign. Because we actually need 
$\mu <0$ for the existence of the DW (see above), this implies that $g<0$,
i.e., {\em attraction} between atoms in BECs (which occurs in $^{7}$Li 
\cite{Li}, as mentioned above, or in $^{85}$Rb \cite{Ru}), is a necessary
condition for the existence of the DW.

The actual shape of the DW differs from the simplest ansatz (\ref{dweq2}) by
the presence of the oscillating part in the wave function at $x\rightarrow
\infty $, see Fig. (\ref{dweq1})(c) below. In the lowest approximation,
treating the OL term in Eq. (\ref{dweq8}) as a small perturbation, one can
find the asymptotic form of the wave function at $x\rightarrow -\infty $ in
the form
\begin{equation}
u_{{\rm asympt}}(x)=\sqrt{\frac{\mu }{g}}\left[ 1-\frac{V_{0}}{2\mu +k^{2}}
\cos \left( kx\right) \right]  \label{corrected}
\end{equation}
[the expression (\ref{dweq5}) was substituted for the amplitude], and the
ansatz (\ref{dweq2}) can be modified accordingly, $u=u_{{\rm asympt}}(x)
\left[ 1-{\rm sgn}\left( x-\xi \right) \right] $.

The divergence of the Lagrangian does not affect the remaining variational
equations, $\partial L/\partial \xi =0$ (cf. the situation in the case of
solitons with nonvanishing tails, where variational equations for the core
of the soliton can also be derived in a divergence-free form \cite{Kaup}).
An eventual result is
\begin{equation}
V_{0}\cos \left( k\xi \right) =\frac{\mu }{4}\frac{2\mu +k^{2}}{\mu +k^{2}}.
\label{cos}
\end{equation}
Equation (\ref{cos}) describes equilibrium between the negative pressure
force acting on the DW from the filled domain (the attraction tends to pull
the atoms in) and the pinning force induced by the OL. The main conclusion
following from Eq. (\ref{cos}) is that a necessary condition for the
existence of the DW solution is that the strength of the OL must exceed a
minimum (threshold) value
\begin{equation}
\left( \left\vert V_{0}\right\vert \right) _{{\rm thr}}=\left\vert \frac{\mu 
}{4}\frac{2\mu +k^{2}}{\mu +k^{2}}\right\vert .  \label{thr}
\end{equation}
We note that the vanishing of the expression (\ref{thr}) in the case
of $2\mu +k^{2}=0$ is a formal feature, produced by the divergence of the
oscillating correction in Eq. (\ref{corrected}) in this case. In fact, this
is a resonant case, and a proper form of the correction can be derived using
well-known methods of the nonlinear-resonance theory. We do not consider
this special case here.

Comparison with numerical results presented in the next section shows that
the exact threshold value is smaller than the one given by Eq. (\ref{thr})
by a factor $\sim 8$. This discrepancy may be explained by the assumption,
implied in the ansatz (\ref{dweq2}), that the DW has zero width; in fact,
numerical results [see Fig. \ref{dfig1}(c) below] demonstrate that it is
narrow indeed, but still its width is not much smaller than $\lambda =2\pi
/k $. The VA can be extended to incorporate a finite width of the DW;
however, the analytical results then become very cumbersome, therefore they
are not displayed here.

\section{Numerical Results}

We will present results of computations for a typical case within the range
of parameters in which the VA predicts the existence of the DW, namely 
for
$g=-3$
and $\mu =-1$. This case can be readily implemented in the experiment,
adjusting the magnitude of the scattering length $a$ (which sets the value
of the normalized nonlinearity parameter $g$) by means of the Feshbach
resonance induced by external magnetic field \cite{fesh} (see also Ref. 
\cite{frm}); for other values of $g$ and $\mu $, the results are quite similar.
With fixed $g$ and $\mu $, in what follows we vary the parameters of the
potential to obtain different types of solutions.

The stationary equation (\ref{dweq8}) was solved by means of a standard
Newton-type algorithm. Once a solution $u_{0}$ to this equation was found,
its linear stability was examined, using a straightforward form of the
perturbed solution, 
\begin{equation}
\psi (x,t)=e^{-i\mu t}\left\{ u_{0}(x)+\left[ \delta a(x)\exp \left(
-i\omega t\right) +\delta b(x)\exp \left( i\omega ^{\star }t\right) \right]
\right\} ,  \label{ab}
\end{equation}
where $^{\star }$ denotes the complex conjugation. The resulting linear
equations for the small perturbations $\delta a$ and $\delta b$ are 
\begin{eqnarray}
\omega \delta a &=&-\left( \delta a\right) ^{\prime \prime }+V(x)\delta
a+2gu_{0}^{2}\delta a-\mu \delta a+gu_{0}^{2}\left( \delta b\right) ^{\star
},  \label{dweq9} \\
\omega^{\ast} \delta b &=&\left( \delta b\right) ^{\prime \prime }-V(x)\delta
b-2gu_{0}^{2}\delta b+\mu \delta b-gu_{0}^{2}\left( \delta a\right) ^{\ast }.
\label{dweq10}
\end{eqnarray}
These can be tackled by means of a matrix eigenvalue solver (using the
finite-difference discretization with a spatial step chosen to be 
$\Delta x=0.2$). The
eigenfrequencies $\omega $ thus found will be shown in the spectral plane 
$(\omega _{r},\omega _{i})$, where the subscripts denote their real and
imaginary parts. The configuration is {\it linearly unstable} if 
there is an eigenfrequency with a non-vanishing imaginary part $\omega
_{i}\neq 0$.

If the stationary solution is found to be unstable, a fourth-order
Runge-Kutta integrator, with time step $\Delta t=0.001$ and no-flux
boundary conditions at the edges of the domain, was used to directly simulate
the development of the instability.

\begin{figure}
\begin{center}
\begin{tabular}{lll}
    ~~~~~~(a) &~~~~~~& ~~~~~~(b) \\
\epsfxsize=8.25cm 
\epsffile{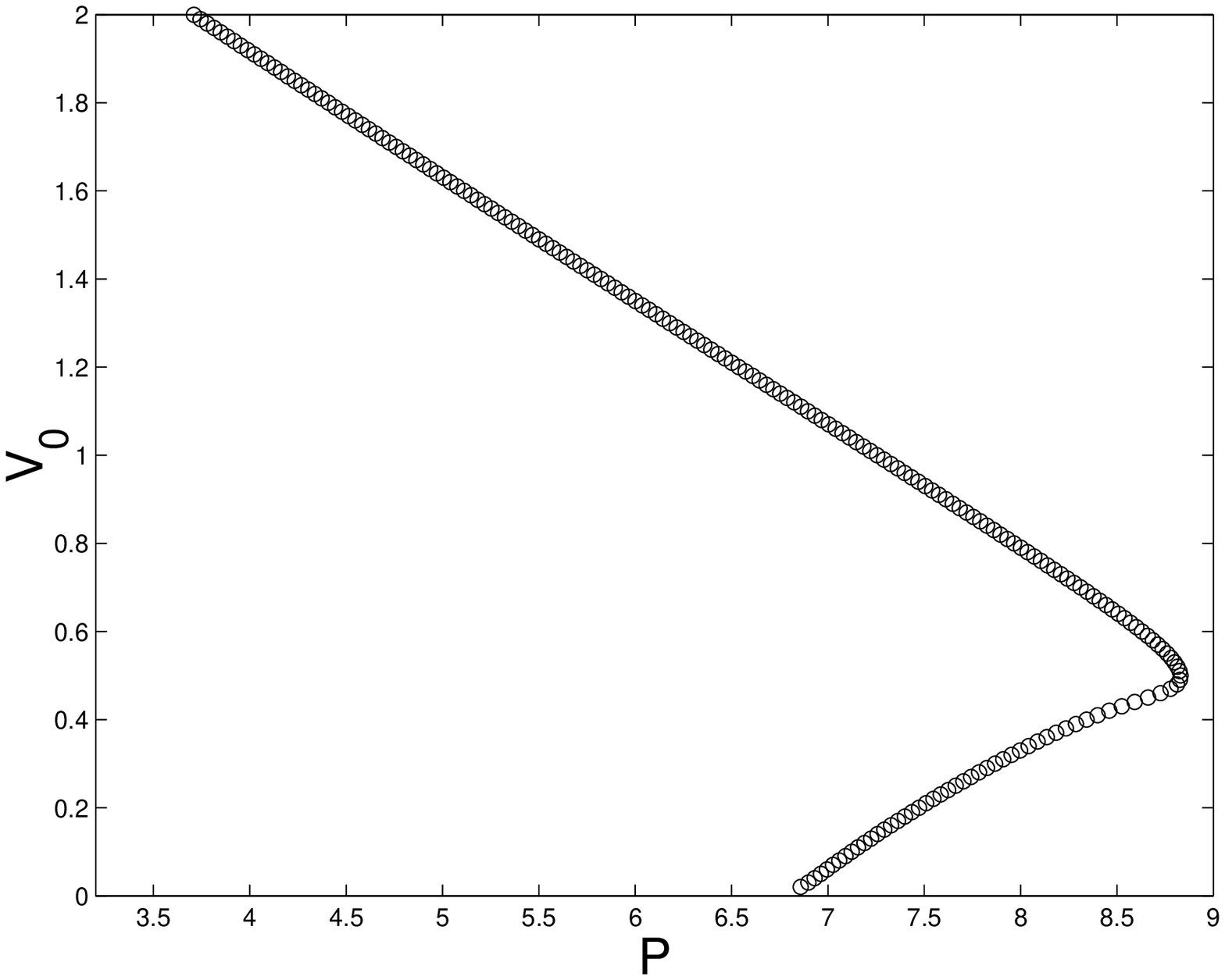} &&
\epsfxsize=8.25cm 
\epsffile{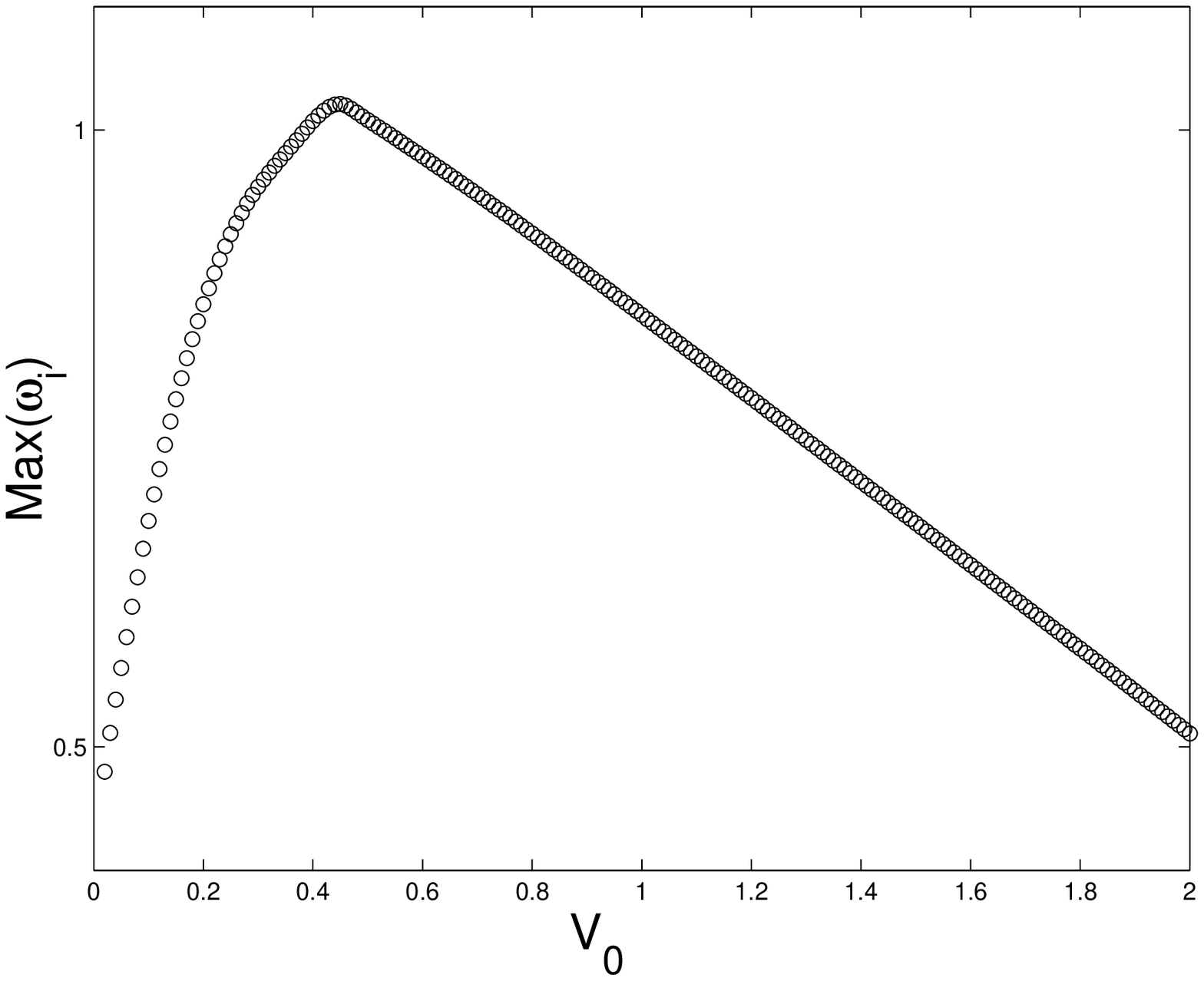} \\[1.0ex]
    ~~~~~~(c) && ~~~~~~(d) \\
\epsfxsize=8.25cm 
\epsffile{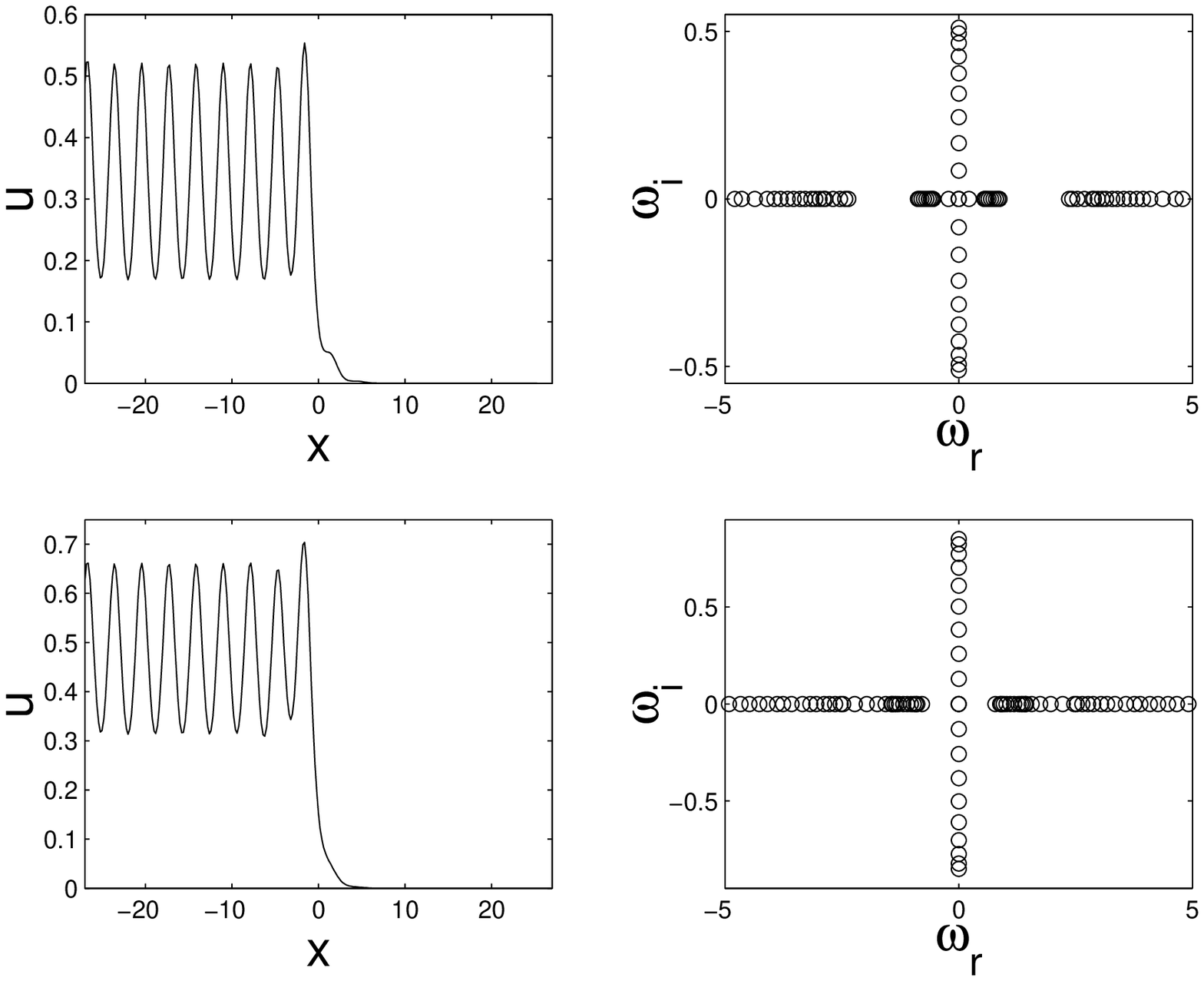} &&
\epsfxsize=8.25cm 
\epsfysize=6.5cm
\epsffile{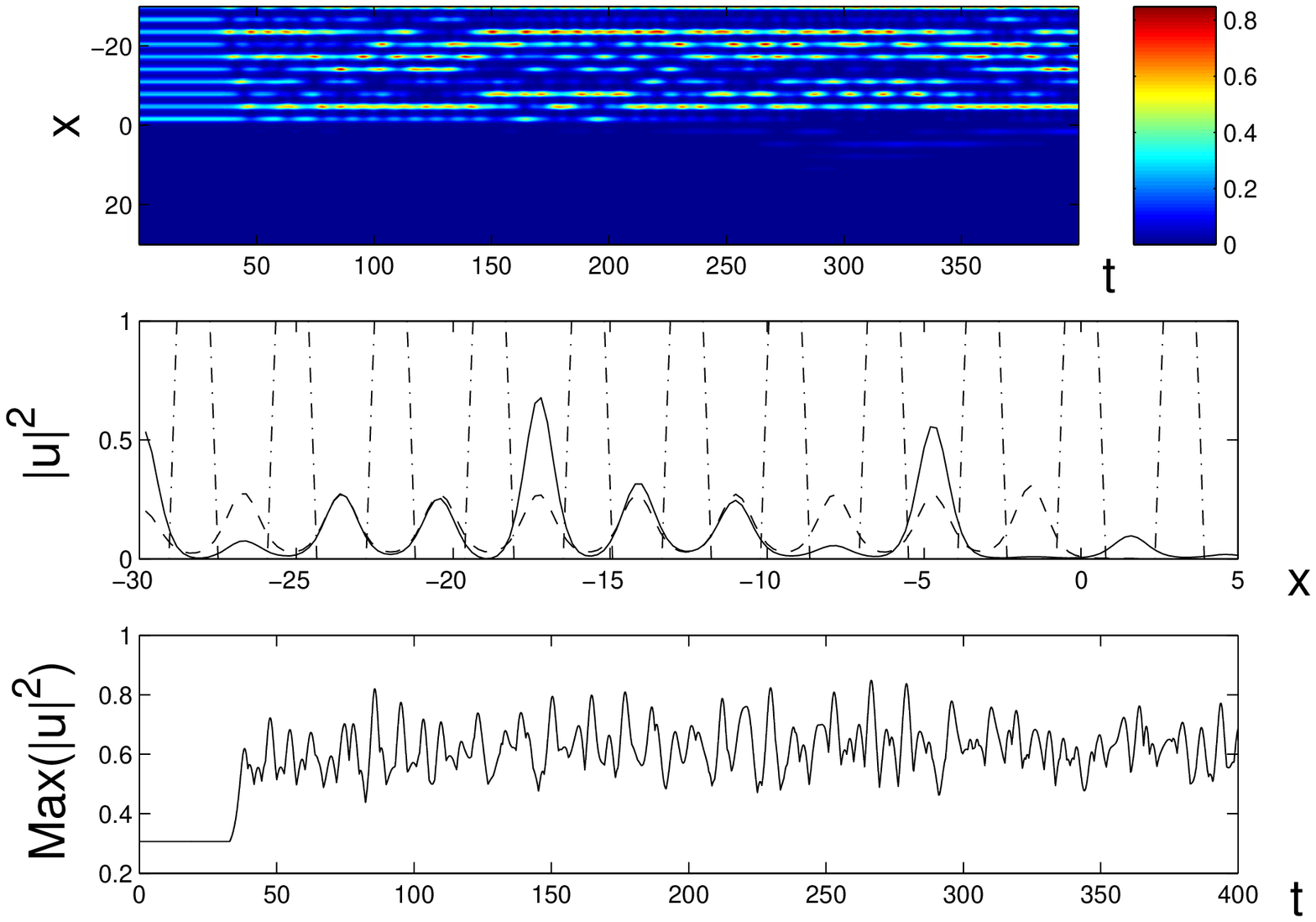} \\
\end{tabular}
\caption{(a) The norm of the domain-wall (DW) solution, 
$P=\protect\int |u|^2 dx$ (alias the normalized number of particles) vs. the
optical-lattice's strength $V_0$ without the external parabolic trap 
($\Omega=0$). The inset explicitly shows that the branch starts at a finite
threshold value of the strength, $(V_0)_{{\rm thr}} \approx 0.02$. (b) The
largest instability growth rate (imaginary part of the eigenfrequency) of
the DW solution vs. $V_0$. This plot shows that the solution is always
unstable. (c) The spatial profile of the solution (left subplots) and the
corresponding spectral plane of the stability eigenvalue $(\protect\omega_r,
\protect\omega_i)$ (right subplots) for $V_0=2$ and $V_0=1$ (top and bottom
subplots, respecively). (d) The space-time contour plot of the density 
$|u|^2 $ showing the evolution of the unstable DW in the case of $V_0=2$ (top
subplot). The middle subplot shows the initial and final configurations
(dashed and solid lines), and the optical-lattice potential (dash-dotted
line). In order to monitor the instability development, the maximum of the
density in the spatial domain is shown, as a function of time, in the bottom
subplot.}
\label{dfig1}
\end{center}
\end{figure}

Numerical results for the stationary DW solutions are displayed in Fig. 
\ref{dfig1}. In this case, we use the OL potential with $k=2$ without the
parabolic trap [$\Omega =0$ in Eq.\ (\ref{pot})], and examine the behavior
of the DW as a function of the potential's strength $V_{0}$. The solutions
indeed exist only if $V_{0}$ exceeds a finite threshold value $\left(
\left\vert V_{0}\right\vert \right) _{{\rm thr}}\approx 0.02$, which is
smaller than the one (\ref{thr}) predicted by the VA in the crude
approximation. A reason for the discrepancy was discussed above, but a
principally important fact is the existence of the finite threshold.

Figure \ref{dfig1} shows that this family of the solutions is unstable at
every value of $V_{0}$, see an explanation below. An example of the
instability development included in Fig. \ref{dfig1} for ($V_{0}=2$) shows
that, on a relatively short time scale ($\sim 30$ in this case), the
instability destroys the DW configuration, leading to an apparently chaotic\
redistribution of the density between different potential wells and gradual
expansion of the condensate into the originally empty domain.

\begin{figure}[tbp]
\begin{center}
\begin{tabular}{lll}
~~~~~~(a) & ~~~~~~ & ~~~~~~(b) \\ 
\epsfxsize=8.25cm \epsffile{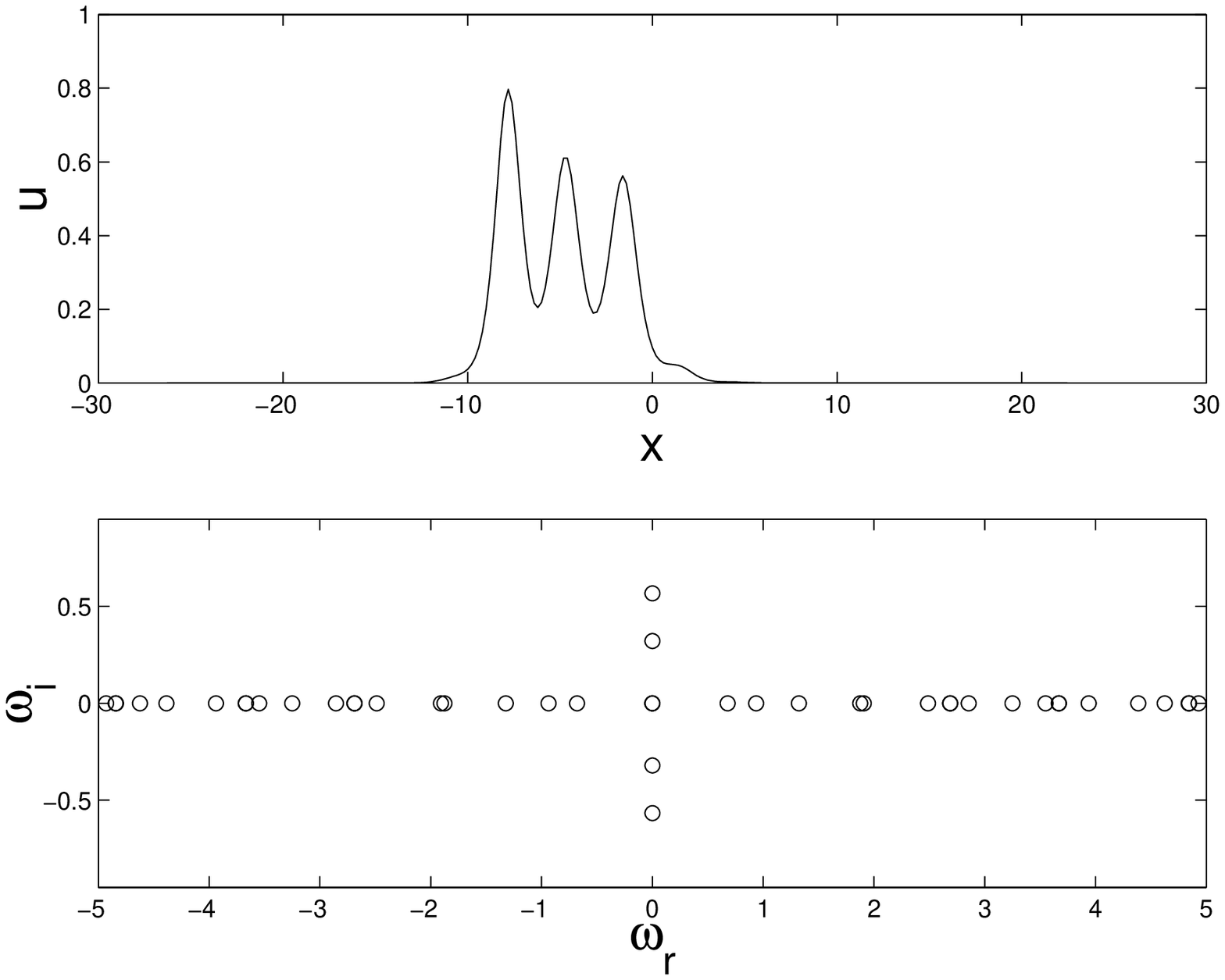} &  & \epsfxsize=8.25cm \epsfysize=6.5cm
\epsffile{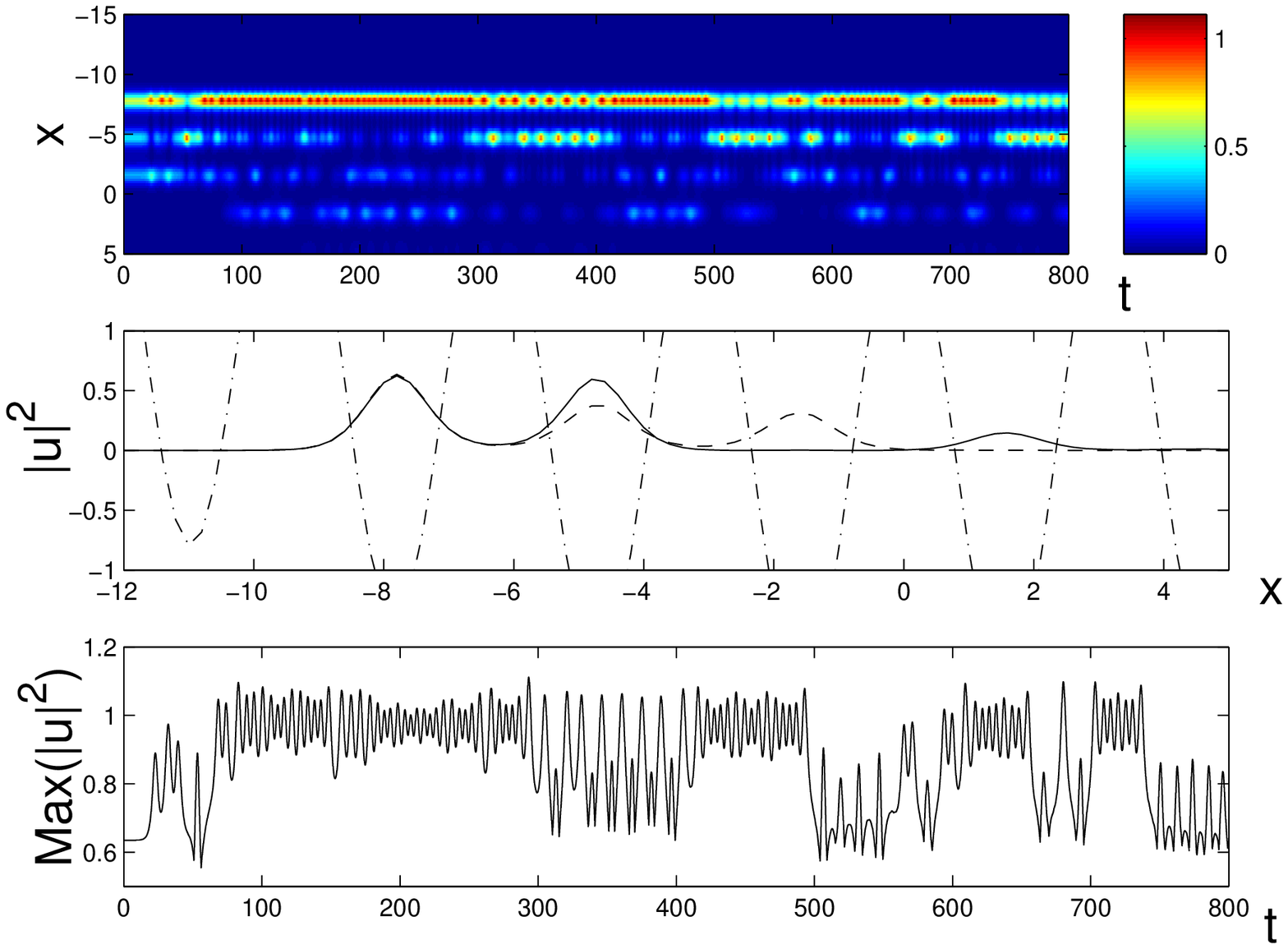} \\[1.0ex] 
~~~~~~(c) &  & ~~~~~~(d) \\ 
\epsfxsize=8.25cm \epsffile{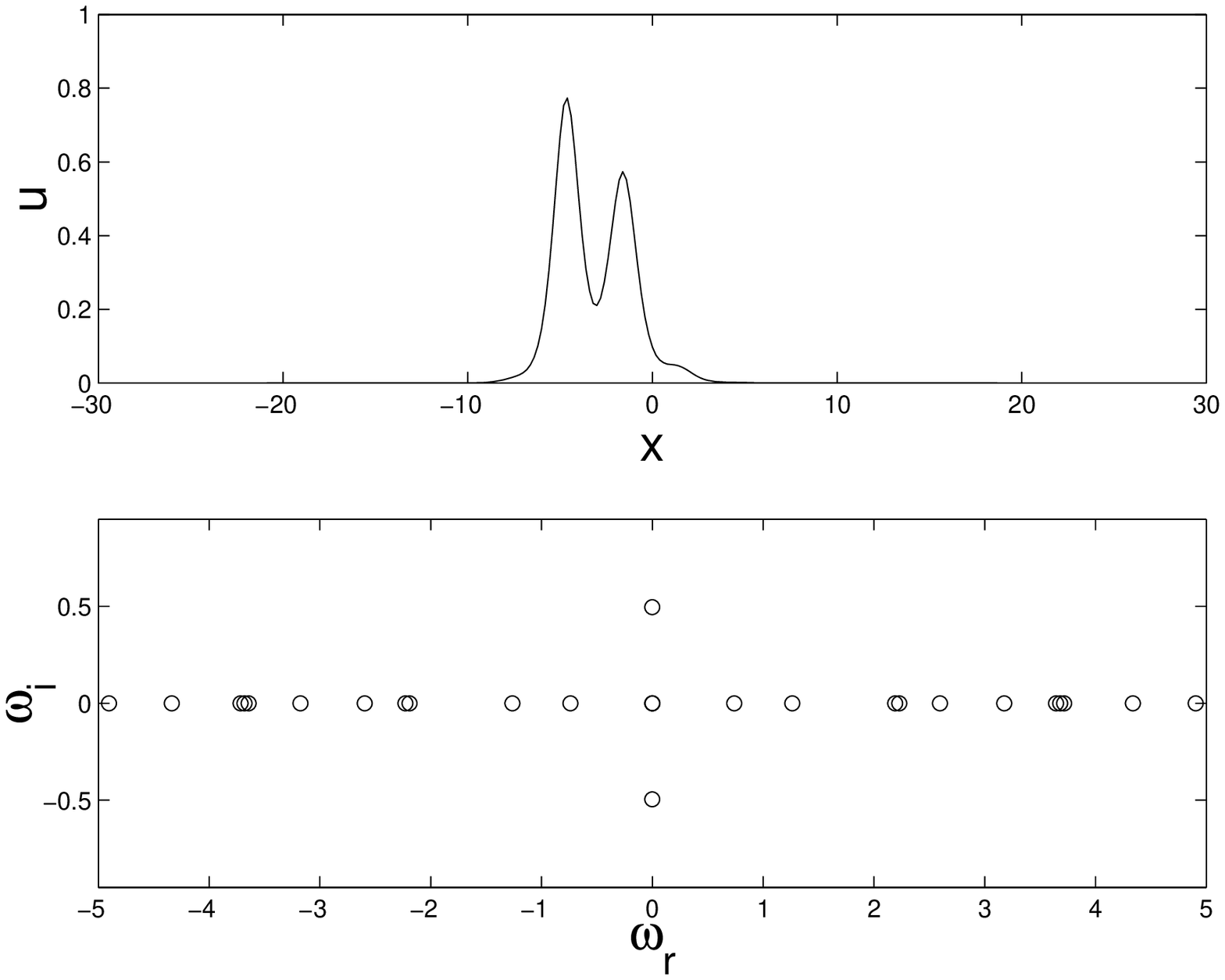} &  & \epsfxsize=8.25cm \epsfysize=6.5cm
\epsffile{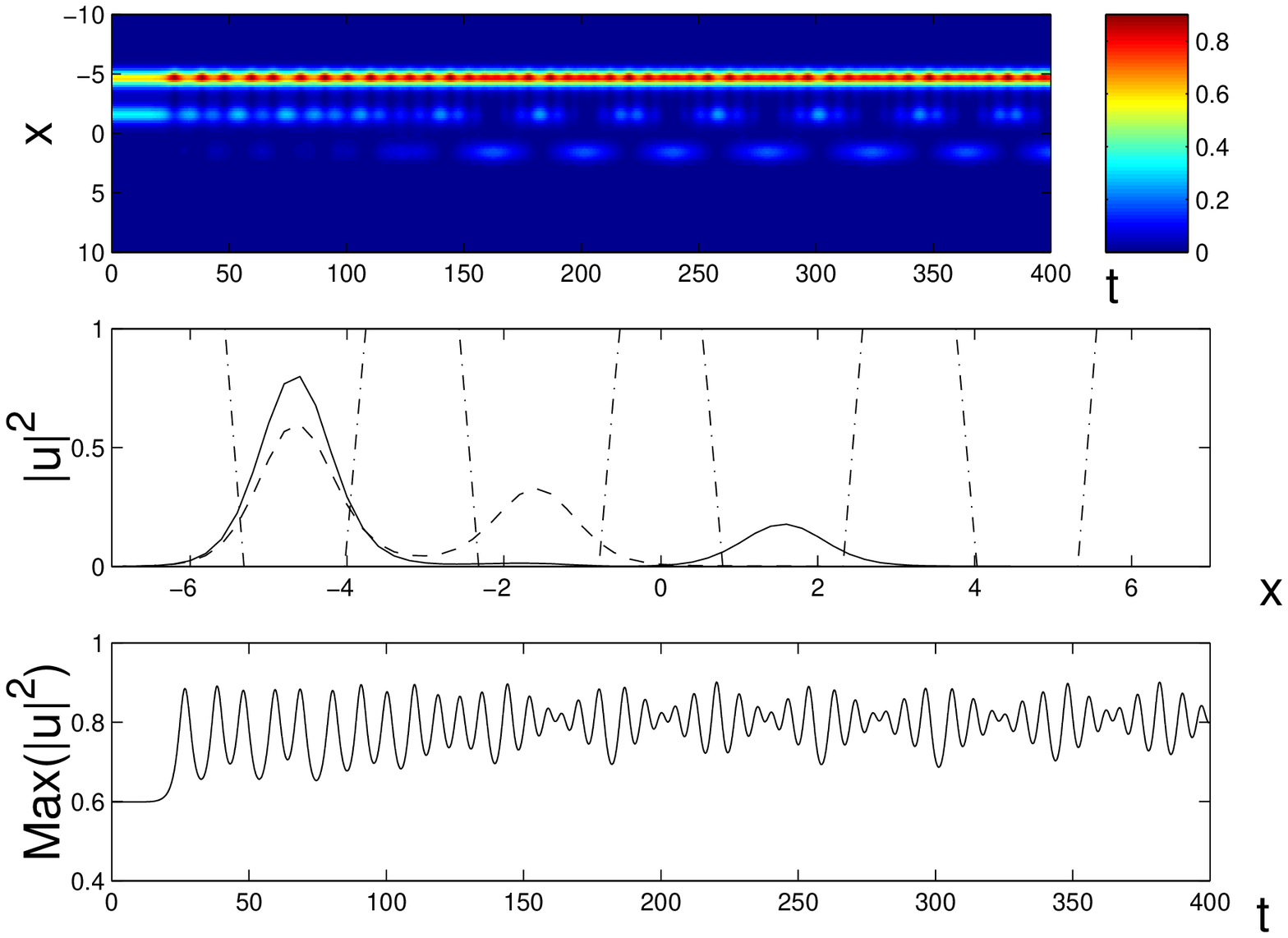} \\ 
&  & 
\end{tabular}
\end{center}
\caption{(a) The profile and spectral plane of a three-humped DW found in
the potential $V(x)=0.01 x^2 + 2 \cos(2x)$. (b) The instability development
of the pattern. Generally, it tends to collect the condensate in a single
potential well. However, there are also time intervals in which the atoms
form two pulses in two adjacent wells (the pulses have opposite signs). (c)
and (d) The same as in (a) and (b), for a double-humped DW in the potential 
$V(x)=0.025 x^2 + 2 \cos(2 x)$.}
\label{dfig2}
\end{figure}

The next step is to consider an effect exerted on DWs by the external
magnetic (parabolic) trapping potential in the system ($\Omega \neq 0$),
which is typically 
an inherent ingredient of experimental BEC settings \cite{review}.
The trap may partly stabilize the DW, as illustrated by Fig.\ \ref{dfig2},
where the combined optical and magnetic potential is imposed, with $\Omega
^{2}=0.01$, $V_{0}=2$ and $k=2$. 
In this case, it is seen that the trap confines the pattern to three humps
in a half of the region $|x|<R_{{\rm TF}}$, where $R_{{\rm TF}}=\sqrt{|\mu
|/\Omega }$ is the Thomas-Fermi radius is ($R_{{\rm TF}}\approx 10$ in this
case; the presence of the DW is evident from the fact that the interval 
$0<x<R_{{\rm TF}}$ is empty). This pattern is still unstable, and its
evolution results in a structure that alternates between a large single-well
pulse and a double-humped one trapped in two adjacent wells.

The stability properties of the three-humped DW configuration considered
above shed light on the origin of the instability of the DWs. In particular,
we observe in the lower part of Fig.\ \ref{dfig2}(a) that this structure has
two unstable eigenfrequencies, which is to be compared with the right part
of Fig.\ \ref{dfig1}(c), where, in the absence of the magnetic trap, {\em
many} unstable eigenfrequencies are present. In this connection, we note
that, by means of the substitution $\delta a=w+v$, $\delta b=w^{\star
}-v^{\star }$, Eqs.\ (\ref{dweq9})-(\ref{dweq10}) can be re-written in the
form of 
\begin{eqnarray}
\omega v &=&-w^{\prime \prime }+V(x)w+3gu_{0}^{2}w-\mu w\equiv -L_{+}w
\label{dweq9a} \\
\omega w &=&-v^{\prime \prime }+V(x)v+gu_{0}^{2}v-\mu v\equiv -L_{-}v.
\label{dweq10a}
\end{eqnarray}
The operator $L_{-}$ has $u_{0}$ as its zero mode, i.e., an eigenstate with 
$\omega =0$. Hence, it follows from Sturm-Liouville theory 
that, since $u_{0}$
does not change its sign, the number of negative eigenvalues of $L_{-}$ is
$n(L_{-})=0$. On the other hand, $u_{0}^{\prime }$ is an $\omega=0$
eigenstate of the
operator $L_{+}$ (in the absence of external potential). 
Again, using Sturm-Liouville theory, it can be deduced (cf. also \cite{todd})
that the number of negative eigenvalues of $L_+$ (if $u_0$ contains 
$N$ separated humps) is 
$n(L_{+})=N$. Consequently, $|n(L_{+})-n(L_{-})|=N$.
One can then infer from the arguments of Refs. 
\cite{ckrtj} and \cite{grillakis} that there must exist $N-1$ unstable real
eigenvalue pairs (imaginary eigenfrequency pairs) in such a case. This
clearly explains the existence of the two unstable eigenvalue pairs in the
case of the three-humped DW.

The above arguments are also supported by panels (c) and (d) of Fig. 
\ref{dfig2}: Here, for a more tightly confining magnetic trap, with $\Omega
^{2}=0.025$ (the OL parameters are again $V_{0}=2$ and $k=2$), the resulting
configuration is a double-humped DW (the larger the magnetic-trap strength 
$\Omega $, the smaller the Thomas-Fermi radius $R_{{\rm TF}}$, hence, the
number of the \textquotedblleft humps\textquotedblright\ decreases). In this
case, $N=2$, and therefore, in agreement with the above arguments, only one
unstable eigenfrequency can be identified in the spectral plane $(\omega
_{r},\omega _{i})$, see the lower part in Fig.\ \ref{dfig2}(c). Figure 
\ref{dfig2}(d) shows that the instability leads to absorption of a large
fraction of the atoms from one ``hump'' by
the other. However, a small fraction of atoms tunnels to the next well. We
have verified that the small pulse generated in this well has parity
opposite to that of the large one on which it abuts. Such opposite-parity
configurations may be {\em stable}, see below.

Previous results for multi-pulse configurations in discrete 
nonlinear-Schr{\"{o}}dinger models with the attractive interaction 
\cite{todd,DKL,TLM,pgk,miw,pjl,njp} have shown that, while patterns composed of
same-parity pulses (i.e., ones with no zero crossings between them) are
always unstable, the so-called {\it twisted localized modes} (TLMs), in
which adjacent pulses are of opposite parity, {\em may be} stable. This can
be seen, in the context of the above consideration, from the fact that if
there are $S$ zero-crossings in the profile $u_{0}$, then $n(L_{-})=S$ and
hence, the Jones-Grillakis criterion \cite{ckrtj}, \cite{grillakis} predicts 
$N-S-1$ real-eigenvalue (imaginary-eigenfrequency) pairs. Thus, if $S=N-1$
(i.e., if adjacent pulses have opposite parities), then the solution is
potentially stable, although other types of instabilities are also known to
occur in this case \cite{TLM,pgk,miw,pjl,njp}, such as the oscillatory
instability (accounted for by a Hamiltonian Hopf bifurcation) \cite{vdm} due
to a negative eigendirection in the energy of small perturbations 
\cite{skryabin,bjorn}.

\begin{figure}
\begin{center}
(a)\hskip5.9cm\null \\
\epsfxsize=8.25cm \centerline{\epsffile{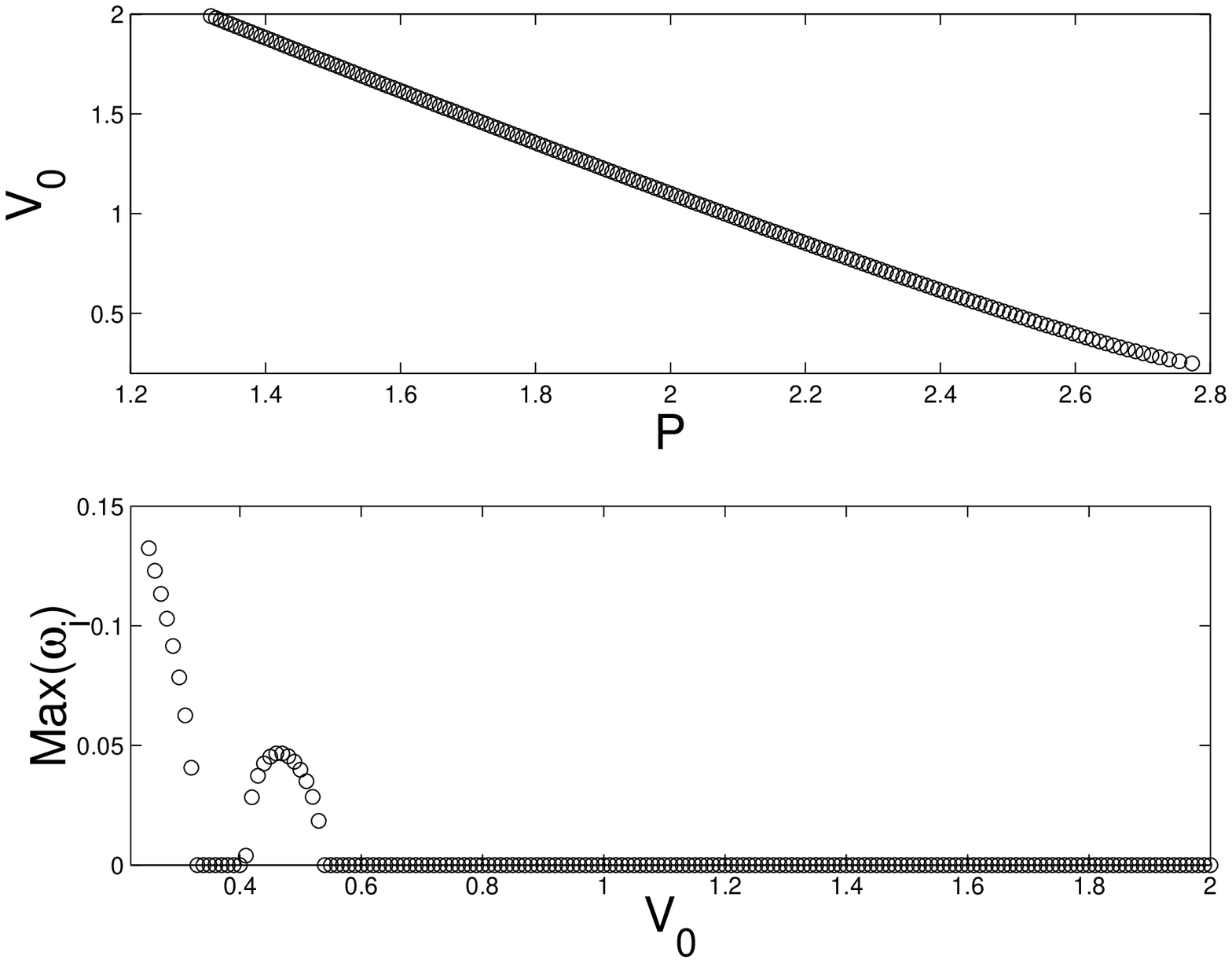}} 
\begin{tabular}{lll}
    ~~~~~~(b) &~~~~~~& ~~~~~~(c) \\
\epsfxsize=8.25cm 
\epsffile{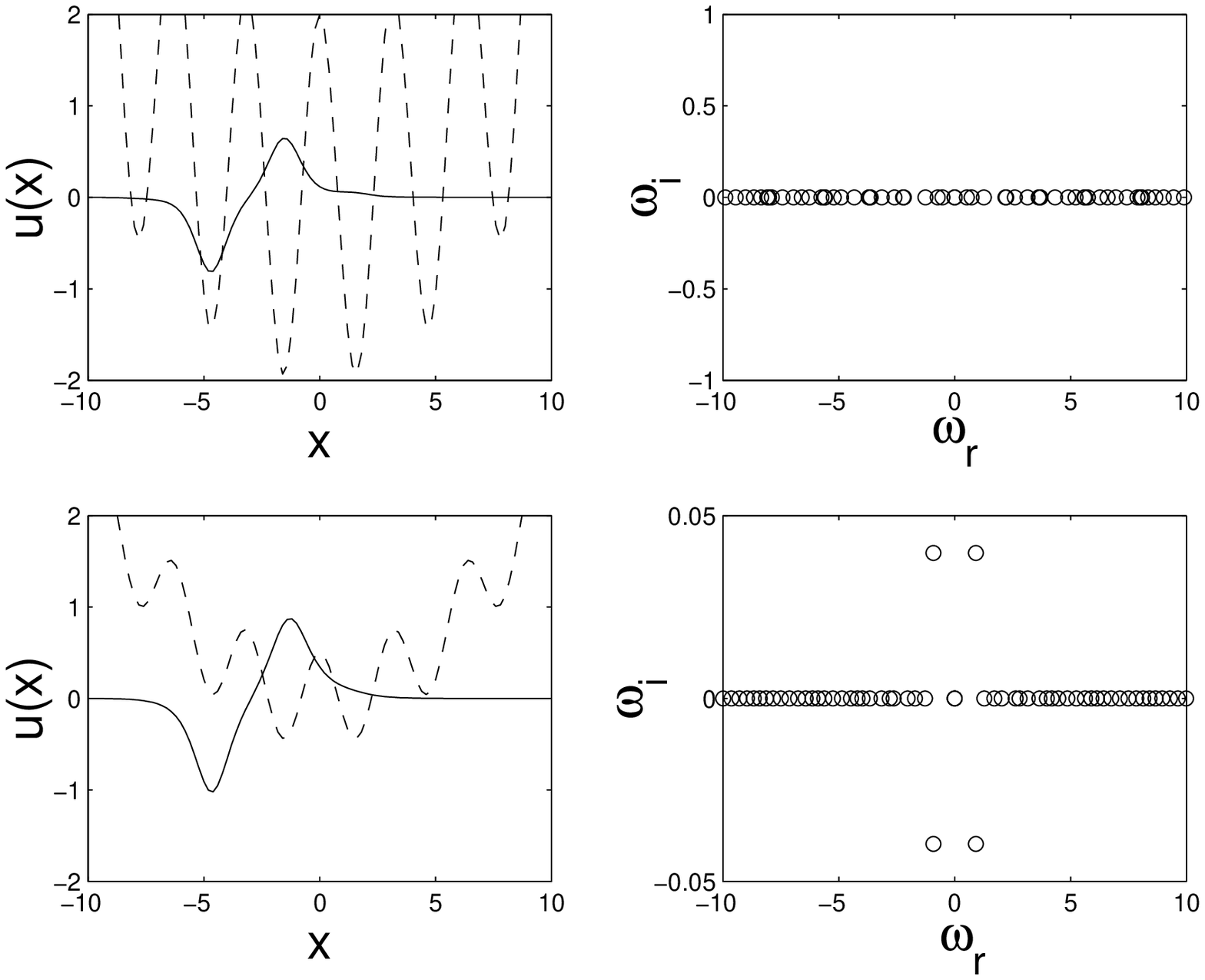} &&
\epsfxsize=8.25cm 
\epsfysize=6.5cm 
\epsffile{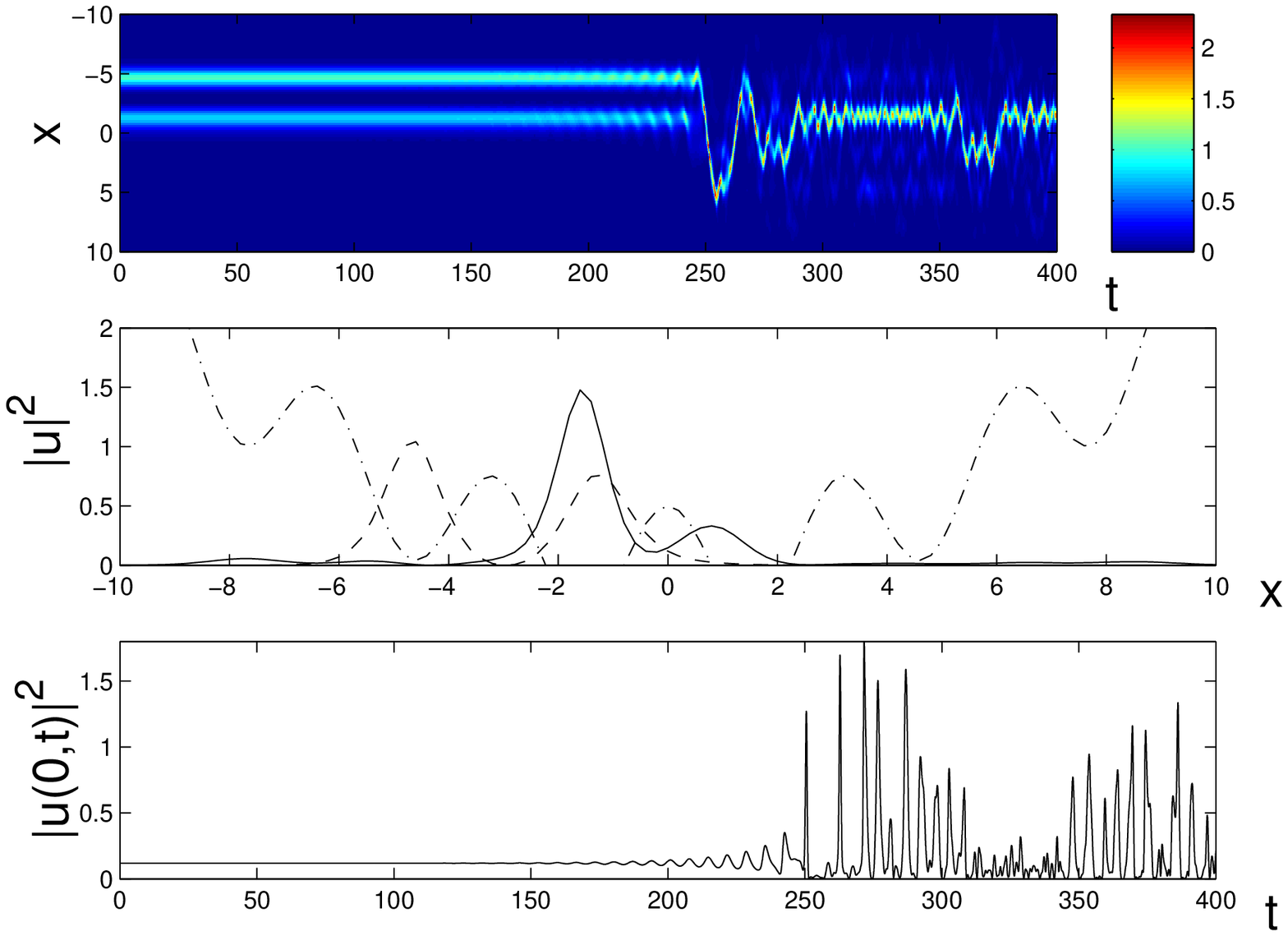} \\
\end{tabular}
\caption{(a) The branch of the twisted-DW solutions. The
bottom subplot of the panel shows the most unstable eigenvalue of small
perturbations around the configuration [the instability, if any, has an
oscillatory character; see panel (b)]. (b) Examples of the twisted DW for 
$V_0=2$ and $V_0=0.5$ (top and bottom subplots) and their 
linear stability spectra
(the dashed line indicates the potential). (c) The evolution of an unstable
twisted DW, which results in transition to an asymmetric single-humped
soliton. Panel (c) is similar to Fig. 1(d), except for the bottom part
illustrating the instability development through the time evolution of the
condensate density at the bottom of the magnetic trap, $|u(0,t)|^2$.}
\label{dfig3}
\end{center}
\end{figure}

Following this argument, we looked for a twisted-DW configuration, which we
were indeed able to identify, as shown in Fig.\ \ref{dfig3}. To examine the
existence and stability of this type of  patterns, we used continuation
in the OL strength $V_{0}$, and found that it exists for $V_{0}\geq 0.25$,
see the top part of Fig.\ \ref{dfig3}(a). Furthermore, it is unstable, due
to the above-mentioned oscillatory instability, in the intervals 
$0.25<V_{0}\leq 0.32$ (i.e., just after it emerges) and $0.41\leq V_{0}\leq
0.53$, while it is {\em stable} otherwise, see the lower part of Fig.
\ref{dfig3}(a). In the case of the instability, we simulated its evolution, as
shown in panel (c) of Fig.\ \ref{dfig3}. It is observed that the instability
cleaves the twisted DW through an oscillatory perturbation [see the top and
bottom subplots of panel (c) in the interval $175\leq t\leq 250$], and
results into oscillations around a configuration with most of the atoms
trapped in a single well. This configuration is actually an asymmetric
soliton (see also below).

\begin{figure}
\begin{center}
\begin{tabular}{lll}
    ~~~~~~(a) &~~~~~~& ~~~~~~(b) \\
\epsfxsize=8.25cm 
\epsffile{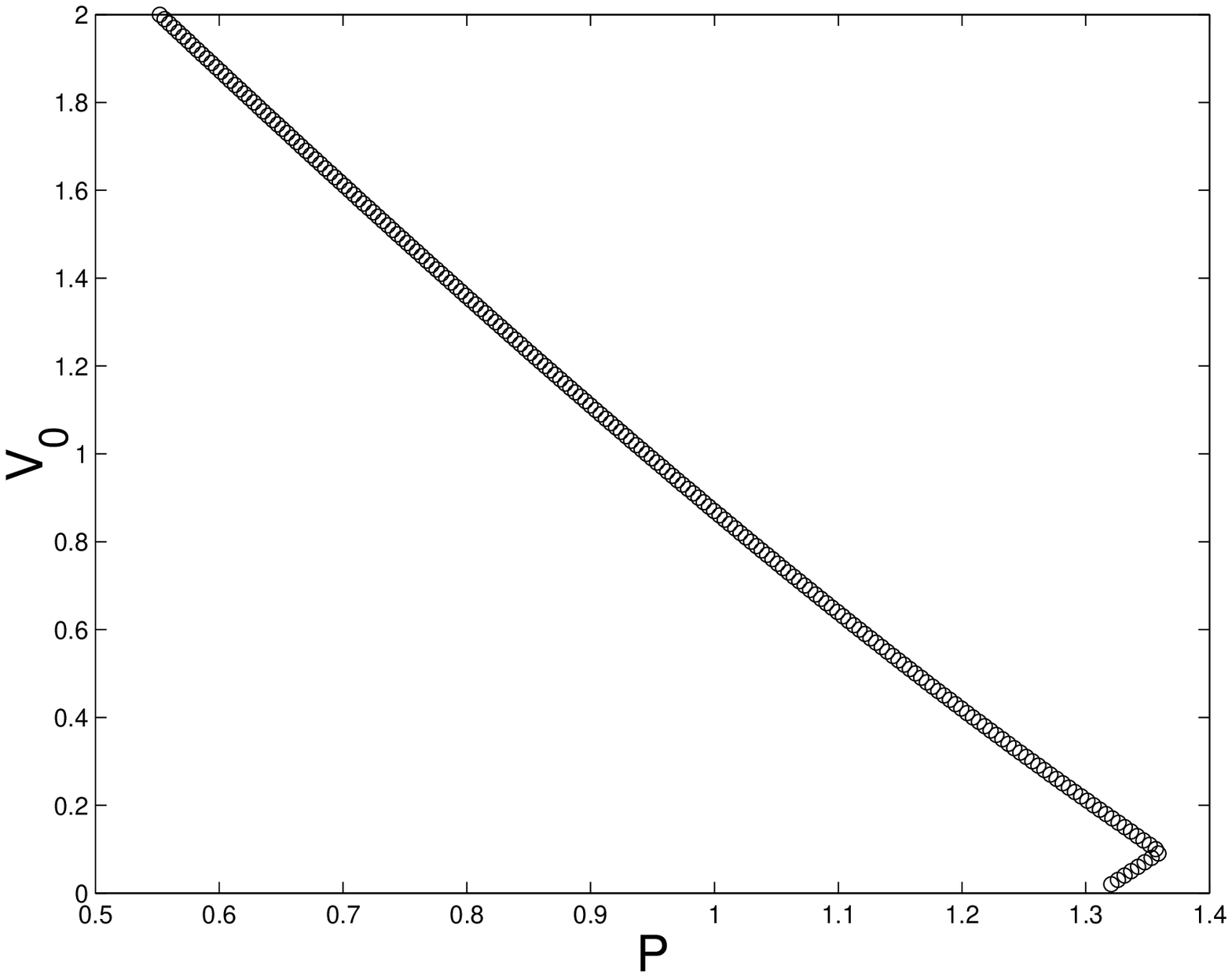} &&
\epsfxsize=8.25cm 
\epsffile{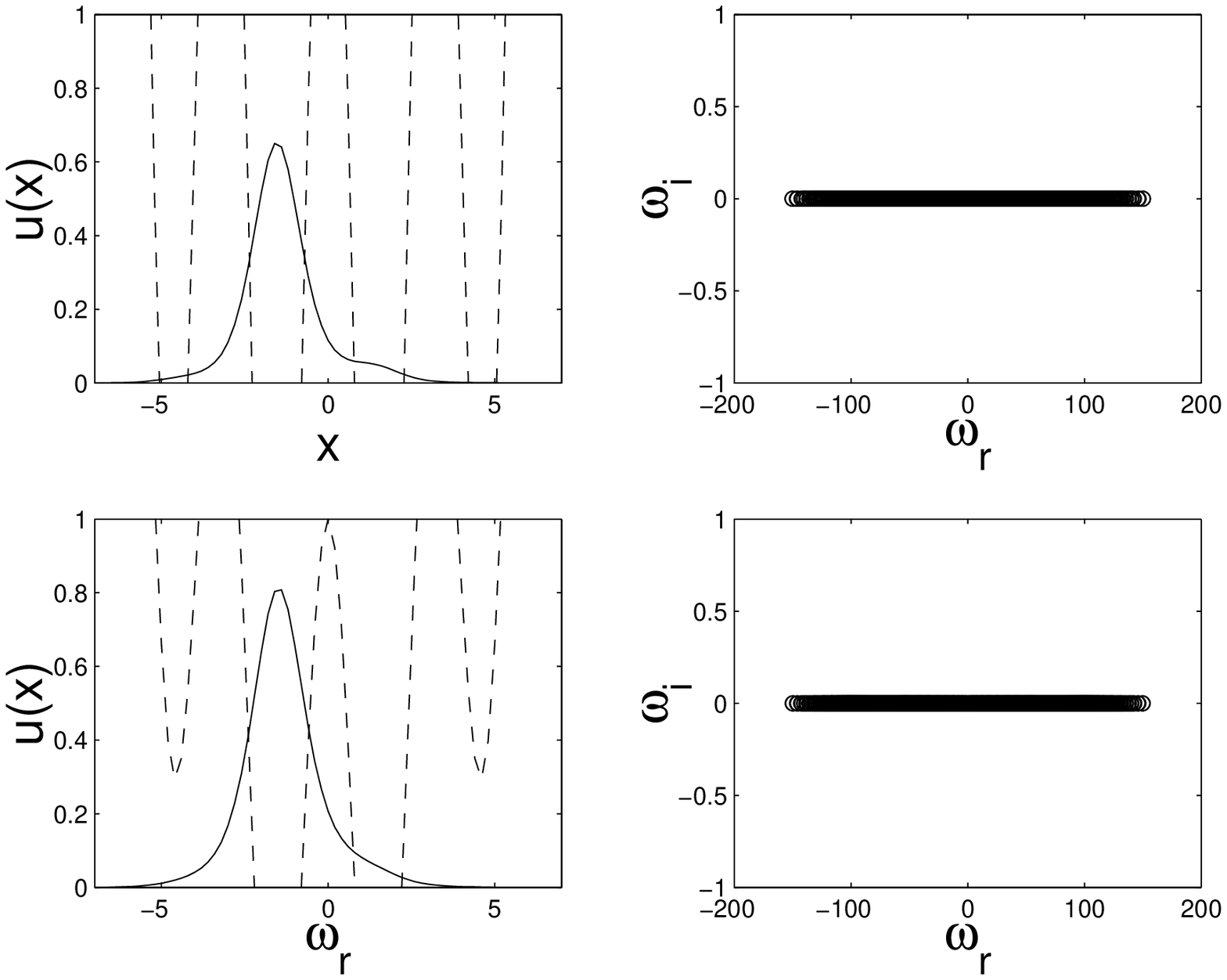} \\
\end{tabular}
\caption{(a) The branch of asymmetric-DW solutions. (b)
Profiles and stability of the solutions for $V_0=2$ and $V_0=1$ (top and
bottom subplots).}
\label{dfig4}
\end{center}
\end{figure}

The latter result, as well as the
natural expectation that, if the Thomas-Fermi
radius $R_{{\rm TF}}$ becomes sufficiently small, a single-humped DW pattern
may emerge, which would be a new type of an asymmetric solitary wave in the
present context, led us to search for a single-humped DW. Such a solution
has been found, and is shown in Fig.\ \ref{dfig4} for different values of 
$V_{0}$ in the potential with $\Omega ^{2}=0.06$ and $k=2$. It is noteworthy
that the solution is stable for {\em all} the examined values of $V_{0}$.
For a small OL strength, the solution degenerates into the regular symmetric
soliton, which is possible in the one-dimensional GP equation with
attraction, and has been observed in the experiment \cite{soliton}.

\begin{figure}[tbp]
\begin{center}
\begin{tabular}{lll}
~~~~~~(a) & ~~~~~~ & ~~~~~~(b) \\ 
\epsfxsize=8.25cm \epsffile{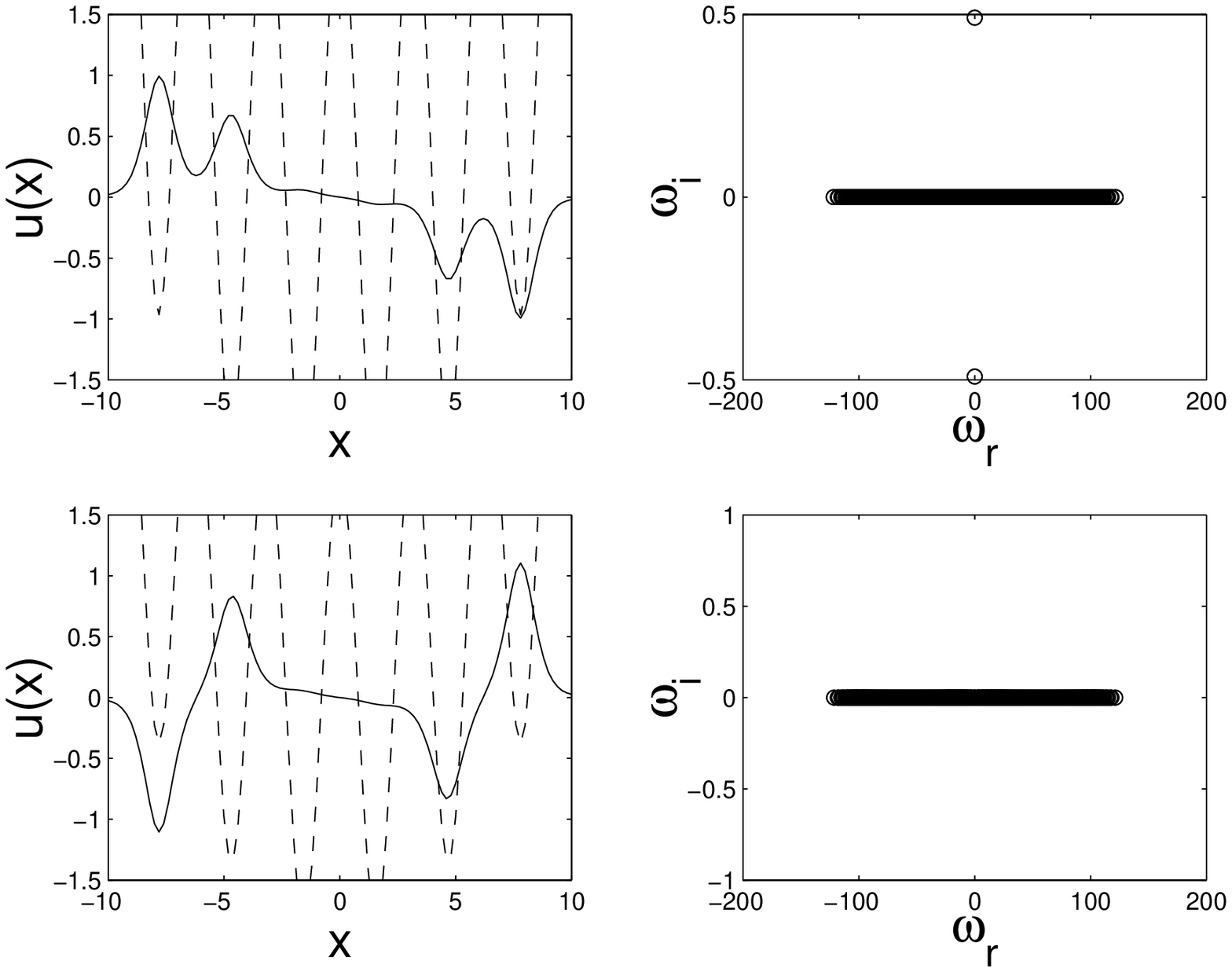} &  & \epsfxsize=8.25cm \epsfysize=6.5cm
\epsffile{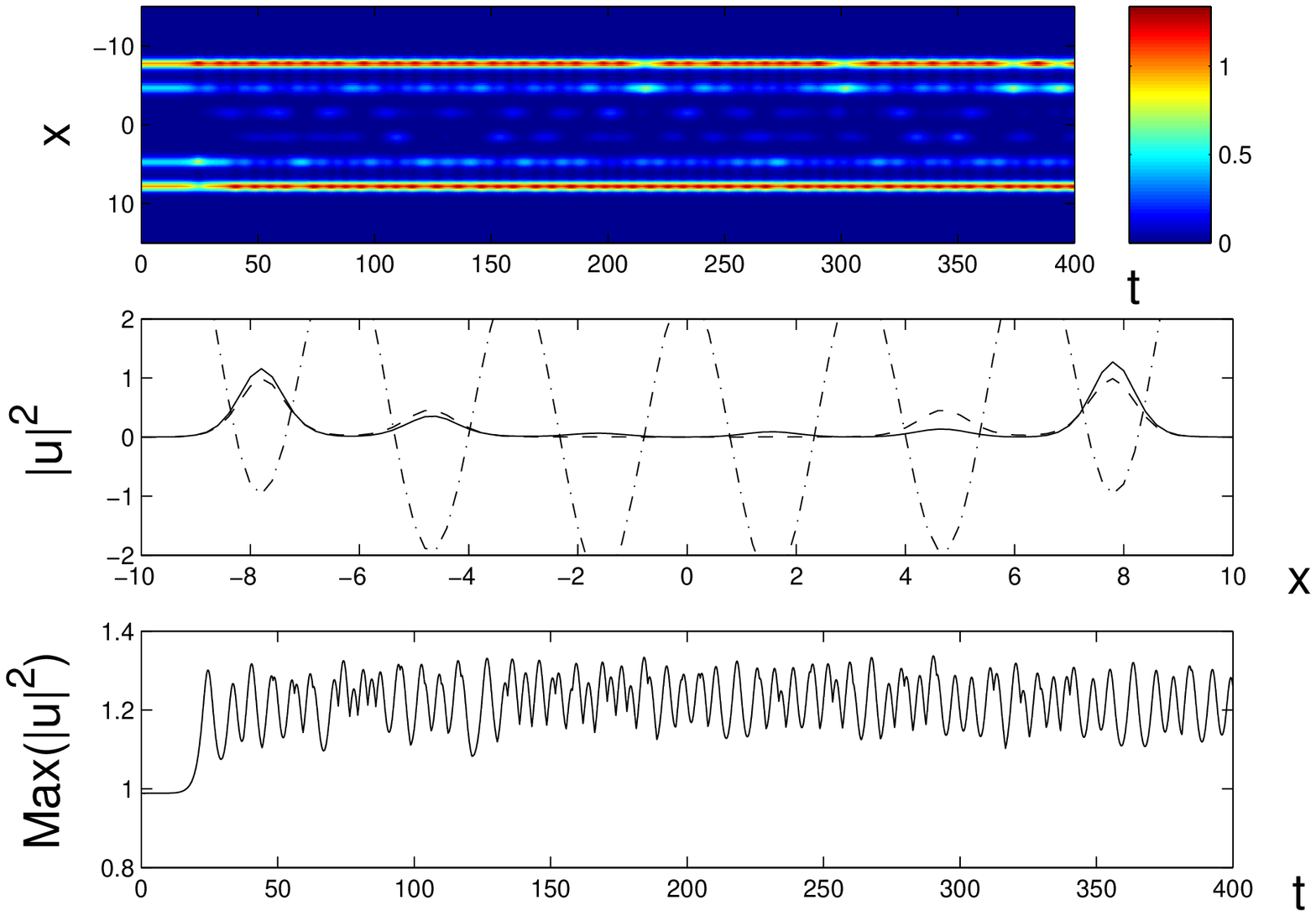} \\ 
&  & 
\end{tabular}
\end{center}
\caption{(a) The bound state of double-humped DWs for $V(x)=0.025 x^2 + 2.5
\cos(2 x)$ (shown by the dashed line) and its stability (right subplot). The
configuration has two nearly identical unstable eigenvalues. The bottom
subplot shows a stable bound state of twisted DWs for $V(x)=0.025 x^2 + 1.9
\cos(2 x)$. (b) The development of the instability of the bound state of
double-humped DWs from the top subplot of (a) leading to the formation of
two larger-amplitude out-of phase pulses. The panel (b) displays 
similar features as  Fig. \protect\ref{dfig1}(d) for the double-humped DW.}
\label{dfig5}
\end{figure}

Finally, we examined possibilities to create composite DWs, constructed as
bound states of the regular or twisted fundamental DWs which were considered
above. Such configurations have been found. Two typical examples, viz.,
bound states of double-humped DWs for $\Omega ^{2}=0.025$, $V_{0}=2.5$ and 
$k=2$, and of twisted DWs for the same $\Omega ^{2}$ and $k$ but for 
$V_{0}=1.9$, are shown in Fig.\ \ref{dfig5}. Naturally, as the former
configuration has one pair of unstable eigenvalues per each double-humped
DW, the composite profile bears two such pairs; similarly, since each single
twisted DW may be stable, their bound state may be stable too (but it may
also be subject to the oscillatory instability shown above). The unstable
evolution of the bound state of the double-humped DWs from the top part of
panel (a) in Fig.\ \ref{dfig5} is shown in panel (b) of the figure. Clearly,
each of the constituent DWs is destroyed in favor of a twisted-DW pattern,
in which the constituent pulses are far separated.

\section{Conclusions}

Domain-wall (DW) solutions appear in a variety of contexts in optics and
matter-wave physics, in both theoretical and experimental studies
(experimentally, DWs have been observed in nonlinear optical fibers). Most
of these settings are inherently two-component ones with immiscible
components, the repulsion between them making it possible to support stable
DWs.

In this work, we have investigated possibilities to create domain walls in 
{\em single-component} models (particularly, in those based on the
Gross-Pitaevskii equation, and relevant to the description of Bose-Einstein
condensates) in the presence of external potentials (OL and parabolic trap).
The same model is relevant to the study 
of spatial solitons in planar waveguides
with transverse modulation of the refractive index. By means of the
analytical approach, based on variational methods, and through 
direct simulations
we have found that DWs are possible only in the model with intrinsic
attraction, and a necessary condition is that the OL strength must exceed a
threshold value. Regular domain walls (with no zero crossings) in the OL
potential are found to be unstable, to which an explanation was given. If
the magnetic trap is introduced, {\it twisted} DWs can be stable, as well as
asymmetric single-pulse soliton solutions. Bound states of elementary DWs
(regular and twisted ones) have also been investigated; they have a chance
to be stable in the latter case.

A natural extension of the present study is to search for multi-dimensional
counterparts of these solutions; these results will be published elsewhere.

P.G.K. gratefully acknowledges support from NSF-DMS-0204585 and from the
Eppley Foundation for Research. Work at Los Alamos is supported by the US
DoE. B.A.M. acknowledges a partial support from the Israel Science
Foundation through a Research-Excellence-Center grant No. 8006/03.


\begin{references}
\bibitem{b24} G.P. Agrawal, {\it Applications of Nonlinear Fiber Optics}
(Academic Press: San Diego, 2001).

\bibitem{b25} Yu.S. Kivshar and G.P. Agrawal, {\it Optical Solitons}
(Academic Press: Amsterdam 2003).

\bibitem{Wab} S. Wabnitz and B. Daino, Phys. Lett. A {\bf 182}, 289 (1993).

\bibitem{Malomed} B. A. Malomed, \newblock Phys. Rev. E {\bf 50}, 1565
(1994).

\bibitem{Haelt} M. Haelterman and A. Sheppard, Phys. Lett. A {\bf 185}, 265
(1994).

\bibitem{MNT} B.A. Malomed, A.A. Nepomnyashchy, and M.I. Tribelsky, Phys.
Rev. A {\bf 42}, 7244 (1990).

\bibitem{Alik} A. Hari and A.A. Nepomnyashchy, Phys. Rev. E {\bf 50}, 1661
(1994); Phys. Rev. E {\bf 61}, 4835 (2000).

\bibitem{Sakaguchi} B.A. Malomed, Phys. Rev. E {\bf 50}, R3310 (1994); H.
Sakaguchi and B.A. Malomed, Physica D {\bf 118}, 250 (1998).

\bibitem{BECdw} M. Trippenbach, K. Goral, K. Rzazewski, B. Malomed, and Y.B.
Band, J. Phys. B: Atom. Mol. Opt. {\bf 33}, 4017 (2000); S. Coen and M.
Haelterman, Phys. Rev. Lett. {\bf 87}, 140401 (2001). P. \"{O}hberg and L.\
Santos, \newblock Phys.\ Rev.\ Lett.\ {\bf 86}, 2918 (2001).

\bibitem{opticalLattice} D.-I. Choi and Q. Niu, Phys. Rev. Lett. {\bf 82},
2022 (1999); A. V. Taichenachev, A. M. Tumaikin, V. I. Yudin, J. Opt. B:
Quant. Semicl. {\bf 1}, 557 (1999); P. Pedri, L. Pitaevskii, S. Stringari,
C. Fort, S. Burger, F. S. Cataliotti, P. Maddaloni, F. Minardi, and M.
Inguscio, Phys. Rev. Lett. {\bf 87}, 220401 (2001); H. Pu, W. Zhang, and P.
Meystre, Phys. Rev. Lett. {\bf 87}, 140405 (2001).

\bibitem{epjd} P. G. Kevrekidis, H. E. Nistazakis, D. J. Frantzeskakis, B.
A. Malomed and R. Carretero-Gonz\'{a}lez, {\it Families of matter-waves in
two-component Bose-Einstein condensates}, Eur. Phys. J. D (in press).

\bibitem{DWexperiment} S. Pitois, G. Millot, and S. Wabnitz, Phys. Rev.
Lett. {\bf 81}, 1409 (1998).

\bibitem{DWarray} J.M. Dudley, F. Gutty, S. Pitois, and G. Millot, IEEE J.
Quant. Electr. {\bf 17}, 587 (2001).

\bibitem{myatt} C. J.\ Myatt, E. A. Burt, R. W. Ghrist, E. A. Cornell, and
C. E. Wieman, \newblock Phys.\ Rev.\ Lett.\ {\bf 78}, 586 (1997).

\bibitem{dsh} D.S.\ Hall, M. R. Matthews, J. R. Ensher, C. E. Wieman, and E.
A. Cornell, \newblock Phys.\ Rev.\ Lett.\ {\bf 81}, 1539 (1998); D.M.\
Stamper-Kurn, M. R. Andrews, A. P. Chikkatur, S. Inouye, H.-J. Miesner, J.
Stenger, and W. Ketterle, \newblock Phys.\ Rev.\ Lett.\ {\bf 80}, 2027
(1998).

\bibitem{KRb} G.\ Modugno, G. Ferrari, G. Roati, R. J. Brecha, A. Simoni,
and M. Inguscio, \newblock Science {\bf 294}, 1320 (2001).

\bibitem{LiCs} M.\ Mudrich, S. Kraft, K. Singer, R. Grimm, A. Mosk, and M.
Weidem{\"{u}}ller, \newblock Phys.\ Rev.\ Lett.\ {\bf 88}, 253001 (2002).

\bibitem{Wang} B. A. Malomed, Z. H. Wang, P. L. Chu, and G. D. Peng, J. Opt.
Soc. Am. B {\bf 16}, 1197 (1999).

\bibitem{GPE1d} V.M.\ P\'{e}rez-Garc\'{\i}a, H.\ Michinel and H.\ Herrero,
Phys.\ Rev.\ A {\bf 57}, 3837 (1998); L.\ Salasnich, A.\ Parola and L.\
Reatto, Phys.\ Rev.\ A {\bf 65}, 043614 (2002); Y.B. Band, I. Towers, and
B.A. Malomed, Phys.\ Rev.\ A 67, 023602 (2003).


\bibitem{review} F.\ Dalfovo, S.\ Giorgini, L.P.\ Pitaevskii, and 
S. Stringari, Rev.\ Mod.\ Phys.\ {\bf 71}, 463 (1999). 

\bibitem{Morsch-Arimondo} O.\ Morsch and E.\ Arimondo, 
\newblock in {\it Dynamics and Thermodynamics of Systems with Long-Range
Interactions}, T. Dauxois, S. Ruffo, E. Arimondo and M. Wilkens (Eds.),
Springer (Berlin 2002), pp. 312-331.

\bibitem{Li} A.J. Moerdijk, W.C. Stwalley, R.G. Hulet, and B.J. Verhaar,
Phys. Rev. Lett. {\bf 72}, 40 (1994).

\bibitem{Ru} J.P. Burke, J.L. Bohn, B.D. Esry, and C.H. Greene, Phys. Rev.
Lett. 80, 2097 (1998).

\bibitem{Kaup} D.J. Kaup and B.A. Malomed, Physica D {\bf 184}, 153 (2003).

\bibitem{fesh} S. Inouye {\it et al.}, Nature (London) {\bf 392}, 151
(1998); E.A. Donley {\it et al.}, Nature (London) {\bf 412}, 295 (2001).

\bibitem{frm} P.G. Kevrekidis, G. Theocharis, D.J. Frantzeskakis, and B.A.
Malomed, Phys.\ Rev.\ Lett.\ {\bf 90}, 230401 (2003).

\bibitem{todd} T. Kapitula, P.G. Kevrekidis and B.A. Malomed, Phys. Rev. 
E {\bf 63}, 036604 (2003).

\bibitem{ckrtj} C.K.R.T. Jones, 
Ergodic Theory and Dynamical Systems {\bf 8}, 119-138 (1988).

\bibitem{grillakis} M. Grillakis, Commun. Pure Appl. Math {\bf 46}, 747
(1988); M. Grillakis, Commun. Pure Appl. Math. {\bf 43}, 299 (1990); M.
Grillakis, J. Shatah and W. Strauss, J. Funct. Anal. {\bf 74}, 160 (1987).

\bibitem{DKL} S. Darmanyan, A. Kobyakov and F. Lederer, \newblock Sov. Phys.
JETP {\bf 86}, 682 (1998).

\bibitem{TLM} P.G. Kevrekidis, A.R. Bishop and K.{\O }. Rasmussen, \newblock
Phys. Rev. E {\bf 63}, 036603 (2001).

\bibitem{pgk} P.G. Kevrekidis, 
{Phys. Rev. E} {\bf 64}, 026610 (2001).


\bibitem{miw} P.G. Kevrekidis and M.I. Weinstein, 
{Math. Comp. Simul.}, {\bf 62}, 65-78 (2003)

\bibitem{pjl} P.J.Y. Louis, E.A. Ostrovskaya, C.M. Savage and Yu.S. Kivshar,
Phys. Rev. A {\bf 67}, 013602 (2003).

\bibitem{njp} P.G. Kevrekidis, D.J. Frantzeskakis, B.A. Malomed, A.R. Bishop
and I.G. Kevrekidis, 
\newblock New J. Phys. {\bf 5}, 64 (2003).

\bibitem{vdm} J.-C. van der Meer, \newblock Nonlinearity {\bf 3}, 1041
(1990).

\bibitem{skryabin} Dmitry V. Skryabin Phys. Rev. E {\bf 64}, 055601 (2001)

\bibitem{bjorn} T. Kapitula, P.G. Kevrekidis and B. Sandstede, Counting
eigenvalues via the Krein signature in infinite dimensional Hamiltonian
systems (preprint).

\bibitem{soliton}
K.E. Strecker, G.B. Partridge, A.G. Truscott, and R.G. Hulet, Nature 
{\bf 417}, 150 (2002); 
L. Khaykovich, F. Schreck, G. Ferrari, T. Bourdel, J. Cubizolles, 
L.D. Carr, Y. Castin, and C. Salomon, Science {\bf 296}, 1290 (2002).


\end{references}
\end{document}